\UseRawInputEncoding
\documentclass[runningheads]{llncs}
\usepackage[T1]{fontenc}
\usepackage{graphicx}
\usepackage{float}
\usepackage{listings}
\usepackage{xcolor}
\usepackage{subcaption}
\usepackage{booktabs}
\begin{document}
%
\title{Annotation of Positive vs Negative User Interactions for Social Sign Prediction}
%
%
\author{Biancamaria Bombino \inst{1,2}\orcidID{0009-0000-5619-5466} \and
Chiara Boldrini \inst{1}\orcidID{0000-0001-5080-8110} \and
Andrea Passarella \inst{1}\orcidID{0000-0002-1694-612X} \and
Marco Conti \inst{1}\orcidID{0000-0003-4097-4064}}
\authorrunning{B. Bombino et al.}
%
\institute{Institute of Informatics and Telematics, CNR, Via G. Moruzzi 1, 56124 Pisa, Italy 
\email{\{b.bombino, c.boldrini, a.passarella, m.conti\}@iit.cnr.it}
\and University of Pisa, Pisa, Italy \\
}
\maketitle              
\begin{abstract}
Inferring the sign of social relationships from online interactions is a fundamental challenge in social network analysis. Existing approaches typically rely on sentiment analysis to label individual interactions as positive or negative, then aggregate these labels to assign a sign to the relationship. However, sentiment analysis captures the valence of the content being discussed rather than the nature of the relational exchange itself, a conflation that can lead to systematic misclassification. In this paper, we propose a methodology that addresses this limitation by leveraging Large Language Models (LLMs) in a zero-shot setting to identify interaction-level relational signals (specifically, personal praise and personal attacks directed at the interlocutor) as more direct indicators of positive and negative social ties. We evaluate four models spanning open-weight and proprietary architectures (Qwen2.5:7b, Gemma2:9b, GPT-4o, GPT-5.4-mini) across three prompt designs of increasing complexity, on two human-annotated datasets of approximately 298 and 340 texts respectively. Results show that zero-shot LLMs achieve good classification performance on both tasks without any task-specific training data, establishing a practical baseline for relational annotation. Performance differs across tasks: attack detection is robust to prompt design and model choice, while praise detection is more sensitive to both, reflecting the greater subjectivity of positive relational gestures. These findings lay the groundwork for integrating LLM-based relational annotation into sign prediction pipelines.


\end{abstract}
\section{Introduction}
\label{sec:intro}
\vspace{-5pt}


Signed networks, graphs in which each edge carries a positive or negative label, have proven to be a powerful abstraction for modelling real-world social systems. The sign attached to a relationship encodes qualitatively distinct information: whether two actors cooperate or compete, trust or distrust, support or oppose one another. Building on Heider's foundational work on cognitive balance and its graph-theoretic formalisation by Cartwright and Harary~\cite{Cartwright1956,Heider1946}, a rich body of research has established that the structure of signed networks follows regularities that unsigned networks cannot capture. Structural balance and status theory have guided empirical analyses of online communities~\cite{Leskovec2010signed}, while signed link prediction has become a benchmark task for graph-learning models~\cite{Beigi2020}. More broadly, the sign of a social tie has direct implications for how we model polarisation, influence, and collective behaviour in both online and offline settings~\cite{Esmailian2014,Kumar2016}.

\vspace{-3pt}
Despite their theoretical and practical importance, signed networks are surprisingly difficult to construct in practice. Signs are unambiguously available only when the platform itself encodes a clear positive/negative semantic, as is the case for review networks such as Epinions, where users explicitly label peers as trusted or distrusted~\cite{Leskovec2010predicting}, or for discussion platforms such as Slashdot, where users tag others as friends or foes. Outside such purpose-built contexts, however, the overwhelming majority of social platforms provide no explicit sign annotations. Researchers must therefore resort to inferring signs from observable interaction traces, a task that raises fundamental ambiguities.

A key difficulty lies in the fact that, on most social platforms, the only observable signal is the \emph{content} of the interactions between users (e.g., the messages they exchange), yet what sign prediction ultimately requires is the valence of the \emph{relationship} itself. These two dimensions are conceptually distinct and easy to conflate: a pair of users who frequently interact around contentious or negative topics may nonetheless maintain a genuinely positive social bond; conversely, two users may share pleasant content while their underlying relationship is one of rivalry or hostility. Tacchi et al.~\cite{tacchi2024keep} propose a methodology to infer the sign of social relationships from Twitter interaction data. Their approach assigns a sign to each relationship based on the fraction of negative interactions between two users, drawing on Gottman's psychological threshold to distinguish positive from negative bonds. To operationalize this, individual interactions are labeled using sentiment analysis of message content, which represents a natural and reasonable proxy. However, sentiment analysis captures the valence of the content being discussed rather than the relational nature of the exchange itself, leaving open the question of whether the resulting sign reflects the relationship between users or merely the topics they tend to discuss. This ambiguity is not specific to Tacchi et al.'s approach: it affects any methodology that relies on interaction-level labeling to infer relationship signs, since the same conflation between content valence and relational valence arises whenever individual interactions are labeled based on their semantic content. Addressing it is therefore a general and open challenge.
Note that this conflation is not a mere implementation detail but a substantive conceptual gap: to date, no systematic methodology exists for inferring the sign of a social relationship while explicitly controlling for the semantic content of the interactions that flow through it.

This paper addresses that gap by posing the following research question: \emph{can we infer the sign of a social relationship from observable textual interactions in a way that captures the nature of the relational exchange rather than the valence of the content being discussed?} 

To answer this question, we propose and evaluate a methodology that leverages large language models (LLMs) in a zero-shot setting. Large language models have demonstrated broad capabilities in natural-language understanding tasks, including sentiment classification and relational inference, without requiring task-specific fine-tuning~\cite{Brown2020}. Rather than relying on content valence as a proxy for relational valence, our approach prompts LLMs to reason explicitly about the nature of the interaction between users, independent of the topic or emotional register of the messages being exchanged. We systematically test a range of prompt designs and demonstrate that, when appropriately prompted, LLMs can effectively distinguish between a positive/negative social relation and a positive/negative content environment within that relation - a distinction that prior approaches have struggled to operationalize.

The main contributions of this work are as follows.
\begin{itemize}
    \item We formally identify and characterize the \emph{relation--content conflation problem} in the context of sign prediction for social networks.
    \item We propose a zero-shot LLM-based methodology that explicitly decouples relational valence from content valence when assigning signs to social links.
    \item We conduct a systematic evaluation of multiple prompt formulations, providing empirical guidance on how to elicit relational reasoning from LLMs.
    \item We demonstrate that appropriately prompted LLMs achieve strong performance in separating positive/negative relations from positive/negative content, establishing a new baseline for interaction-level relational annotation that can be integrated into any sign prediction pipeline that aggregates interaction labels into relationship signs.
\end{itemize}

The remainder of this paper is organized as follows. Section~\ref{sec:related} reviews related work on signed networks, sign prediction, and the use of LLMs for social-network analysis. Sections~\ref{sec:methodology} and ~\ref{sec:datasetecc} describe our methodology and experimental settings. Section~\ref{sec:results} presents our experimental results. Section~\ref{sec:discussion} discusses implications and limitations, and Section~\ref{sec:conclusion} concludes the paper.

\vspace{-10pt}
\section{Related work}
\label{sec:related}
\vspace{-5pt}
 
Signed networks extend traditional graph models by associating each edge with a polarity, distinguishing positive relationships, such as trust and homophily, from negative ones, such as distrust and antagonism~\cite{Tang_2016}. This additional information has proven useful in a variety of tasks, including community detection~\cite{Traag_2009} and information diffusion~\cite{Ferrara_2015}.
However, explicit signed networks are relatively rare in real-world platforms. When link polarity is directly observable, negative ties typically account for only a small fraction of the total than the positive ones, often between 15--23\%~\cite{Leskovec2010signed}. This imbalance is commonly attributed to social pressures that discourage users from expressing negative relationships openly~\cite{Coleman_1988}.
As a consequence, a significant line of research has focused on inferring the sign of relationships in networks where this information is not explicitly available. Many existing approaches rely on structural properties of the graph, leveraging topological features or learning-based models trained on annotated datasets~\cite{Javari_2014,Ye_2013}. While effective in some settings, these methods adopt a top-down perspective and do not directly consider the semantic content of user interactions.

One of the earlier approaches to the problem of inferring the sign of a relationship from user interactions is proposed by Hassan et al.~\cite{Hassan_2012}, who trained a Support Vector Machine (SVM) on a manually annotated dataset of relationships in discussion forums. The model uses both user-level and interaction-level features and achieves an accuracy of 0.835 on a subset of annotated data. However, this approach is not directly applicable to platforms such as Twitter, where interactions are significantly shorter and less structured. Furthermore, the lack of publicly available ground truth data for relationships limits its applicability in real-world settings.

An alternative direction is to infer relationship polarity from the sentiment expressed in individual interactions. Sentiment analysis at the level of single exchanges is a well-established problem~\cite{Liu_2012}, allowing reliable labeling of individual messages. Since relationships emerge from repeated interactions, this provides a natural basis for inferring their overall sign. However, extending sentiment signals from individual exchanges to entire relationships remains a challenging problem and has received comparatively limited attention.


Tacchi et al.~\cite{tacchi2024keep} propose a principled methodology to infer the sign of social relationships from Twitter interaction data. They assign a sign to each relationship based on the fraction of negative interactions between two users, drawing on Gottman's psychological threshold (a relationship is classified as negative if more than one in five interactions is negative), and positive otherwise. To operationalize this, individual interactions are labeled via sentiment analysis, which represents a natural and reasonable proxy given that message content is the only directly observable signal on such platforms. However, sentiment analysis captures the valence of the content being discussed rather than the relational nature of the exchange itself, leaving open the question of whether the resulting sign reflects the bond between users or merely the topics they tend to discuss.
In this work, we address this limitation by proposing an approach specifically designed for social textual interactions. Rather than relying on sentiment as a proxy, we leverage Large Language Models to capture the nature of interactions between users, explicitly focusing on relational signals independently of the content or emotional tone of the exchanged messages.

\vspace{-10pt}
\section{Methodology}
\label{sec:methodology}
\vspace{-5pt}

This section describes the methodological framework we propose to infer the sign of social relationships from textual interactions, without relying on content sentiment as a proxy for relational valence.

\vspace{-10pt}
\subsection{Task formulation}
\vspace{-5pt}

As discussed in Section~\ref{sec:intro}, the core challenge in sign prediction from textual data is that the only observable signal (message content) does not directly reveal the nature of the relationship between users. Rather than reading content sentiment as a proxy for relational valence, we propose to identify specific interactional signals that are more directly indicative of the sign of a social tie: \emph{personal praise} and \emph{personal attacks}.

Personal praise refers to expressions that convey a positive disposition toward the interlocutor as a person, independently of the topic being discussed. Personal attacks refer to expressions of hostility or negativity directed at the interlocutor as a person, again independently of the content of the exchange. Crucially, both dimensions are defined at the level of the \emph{relational gesture} rather than the \emph{content valence}: a message discussing a negative topic can still constitute praise, and a message with neutral content can still constitute an attack.

We therefore formulate sign prediction as two independent binary classification tasks: given a textual interaction, determine whether it constitutes (i) a personal praise directed at the interlocutor, and (ii) a personal attack directed at the interlocutor. The sign of the relationship between two users can then be inferred by aggregating these interaction-level labels across all exchanges between them. This makes our methodology applicable whenever interaction-level signals are used to derive relationship-level signs — a general paradigm instantiated by appraches such as Hassan et al.~\cite{Hassan_2012} and Tacchi et al.~\cite{tacchi2024keep}, but which extends to any approach that aggregates individual interaction labels into a relationship polarity.

\vspace{-10pt}
\subsection{LLM-based annotation}
\vspace{-5pt}

We propose to operationalize the two classification tasks using Large Language Models in a zero-shot setting. This choice is motivated by two considerations. First, LLMs have demonstrated strong natural language understanding capabilities across a wide range of tasks without requiring task-specific fine-tuning~\cite{Brown2020}, making them well-suited to capture the subtle linguistic cues that distinguish relational gestures from content sentiment. Second, and more importantly, LLMs can be explicitly prompted to reason about the nature of an interaction (whether it constitutes a gesture toward the interlocutor) rather than about the emotional tone of the content, which is what sentiment analysis captures. This makes them a natural fit for our task formulation.
We evaluate a range of prompt designs to assess how the level of instruction detail influences annotation quality. Specifically, we consider three prompt variants of increasing complexity: a \emph{minimal} prompt providing only the core task definition, an \emph{intermediate} prompt adding clarifications about the directionality of the gesture (i.e., toward the interlocutor vs. toward a third party), and a \emph{structured} prompt providing explicit positive and negative examples of each class. This variation allows us to empirically assess whether more detailed instructions improve the model's ability to elicit relational rather than content-level reasoning, or whether they introduce noise that reduces generalization. All prompts can be found in Appendix~\ref{app:prompts}.

\vspace{-10pt}
\subsection{Testing ensembles of LLMs}
\label{sec:methodology_ensemble}
\vspace{-5pt}

As an exploratory extension, we investigate whether annotation quality can be further improved by combining multiple LLMs rather than relying on a single model. We propose a cascading ensemble strategy that exploits complementary error patterns across models.
The cascade operates as follows. A primary model is selected based on its reliability on one of the two classes (specifically, the class for which it exhibits the lowest error rate). This model classifies the entire dataset, and, for the remaining instances (those belonging to the class where the primary model is less confident), a second model is applied. The second model is selected on the basis of two criteria: complementarity in performance with respect to the primary model, and low correlation in error patterns on the relevant subset. This process can be iterated, introducing additional models for progressively more uncertain residual cases. Error correlation matrices computed on held-out predictions are used to guide model selection at each stage.
Whether this strategy yields consistent improvements over the best single-model configuration is an empirical question that we investigate in Section~\ref{sec:results}. 

\vspace{-10pt}
\section{Experiment settings}
\label{sec:datasetecc}
\vspace{-5pt}

This section describes the concrete instantiation of the methodology introduced in Section~\ref{sec:methodology}, covering the datasets constructed to evaluate the two classification tasks, the specific models and configurations used, and the evaluation procedure.

\vspace{-10pt}
\subsection{Datasets}
\vspace{-5pt}

We created two different datasets, one for each semantic dimension introduced in Section~\ref{sec:methodology}: personal praise and personal attacks. In both cases, human annotations are used to establish ground truth labels via majority voting.

\vspace{-10pt}
\subsubsection{Praise dataset}
First, we built the dataset of praise comments using the publicly available GoEmotions corpus.\footnote{\url{https://www.kaggle.com/datasets/debarshichanda/goemotions}}
This dataset is composed of 58,000 comments extracted from Reddit, with human annotations for 27 emotional categories or the neutral state. Although this corpus may be present in the pre-training data of some of the evaluated models, this does not directly threaten the validity of our evaluation. The original GoEmotions labels are not used in our study: instead, we select a subset of comments and re-annotate them from scratch according to our own annotation scheme, asking human annotators to label each comment as personal praise or not. Since the labels the models are asked to predict are entirely different from any annotation the corpus may have carried during pre-training, the risk of label leakage is limited.
Specifically, we picked approximately 500 comments characterized by positive sentiment in which the emotion \textit{admiration} was detected, as these are likely to contain expressions of praise. 
Each comment in this subset was then labeled by 7 human annotators as \textit{praise (1)} or \textit{not praise (0)}, and the final ground truth labels were obtained via majority voting. 
To avoid an imbalance between classes, we constructed a balanced dataset by selecting an equal number of instances from each class based on the majority labels. 
Consequently, the final dataset consists of 298 comments: 149 \textit{praise} and 149 \textit{non-praise}.

\vspace{-10pt}
\subsubsection{Personal Attacks dataset}
In the case of personal attacks, no suitable existing corpus was available: to the best of our knowledge, there is no publicly available dataset of social media interactions annotated (or even collected) specifically for the personal attack vs. non-attack distinction as we define it here. We therefore constructed the dataset from scratch.
The dataset was constructed by combining data downloaded from two different sources: Wikipedia talk pages, which represent discussion pages and exchanges of criticism among editors, and Reddit. From the latter social platform, comments were extracted from specific subreddits such as \textit{"roastme"} to ensure we obtained real personal attacks. This choice was made to capture both criticism and explicit hostility.  
From these sources, a total of approximately 600 comments dating from 2024 to 2026 and filtered by negative sentiment were collected and subsequently annotated by 5 human annotators. Each text was labeled as a \textit{personal attack (1)} or \textit{not attack (0)}, and the final labels representing the ground truth were again determined by majority vote. 
As in the previous case, the dataset was finally balanced to ensure an even distribution of the two classes. 
In conclusion, the final negative corpus consists of 340 comments: 170 attacks and 170 non-attack.

\vspace{-15pt}
\subsubsection{Dataset comparison in annotation complexity}

To characterize the difficulty of the two annotation tasks, we analyzed the distribution of annotator agreement in both datasets. Borderline cases, where only a bare majority agrees on the label, reveal the inherent subjectivity of each task and provide useful context for interpreting model performance.

The distribution of agreement reveals a slight difference between the two tasks. In the case of personal attacks, the majority of cases (57.35\% - 195/340) show complete consensus among annotators, indicating that such interactions are often clearly identifiable. Only a small percentage of cases (14.41\% - 49/340) fall into the borderline category.
In contrast, the dataset regarding praise shows a more dispersed agreement pattern. Unanimity is achieved in only 20.80\% (62/298) of cases, while a substantial portion of the annotations is distributed across intermediate levels (5/7 and 6/7). These correspond to 24.49\% (73/298) and 37.58\% (112/298) of cases respectively. 

This may suggest that expressions of praise are interpreted less consistently, with humans differing more frequently in their judgments.  It should be noted, however, that this difference may reflect characteristics specific to the corpora we constructed rather than a general property of the two classes.
Overall, although both tasks include ambiguous cases, the agreement patterns indicate that, at least in our datasets, hostility tends to be a bit more explicitly recognizable, while acclaim is expressed in more varied and context-dependent ways, leading to greater variability in annotation. The observed agreement patterns will serve as a useful reference when interpreting model error distributions in Section~\ref{sec:results}.


\vspace{-10pt}
\subsection{Models used and their configuration}
\vspace{-5pt}
We evaluate four models spanning two categories: GPT-4o and GPT-5.4-mini as proprietary black-box models, and Qwen2.5:7b and Gemma2:9b as open-weight alternatives. 
This selection enables comparison across different types of architectures, training paradigms, sizes, and levels of accessibility.
Each model is evaluated under the three prompt variants introduced in Section~\ref{sec:methodology} (minimal, intermediate, and structured) applied independently to both tasks, for a total of 24 experimental configurations.

To ensure consistency and reproducibility, deterministic decoding was used with the temperature set to 0 for all models, in order to eliminate randomness and creativity from the generation process.
Under this setting, model outputs are expected to be stable across runs. But to empirically verify this behavior, we additionally repeated each experiments 3-5 times and observed no variation in the results.
Furthermore, when the reasoning is not required to verify the process, the maximum number of tokens generated is limited (e.g., 10-- 50 tokens depending on the model) to produce concise outputs, as the task requires only a binary label. This constraint also reduces computational overhead and improves inference efficiency. Model outputs were requested in a structured JSON format (e.g., \texttt{{"label": 0}}), with a post-processing step to extract valid JSON from outputs that deviated from the expected format.



\vspace{-10pt}
\subsection{Performance metrics}
\vspace{-5pt}

Model predictions are evaluated against human-annotated ground truth using standard binary classification metrics: accuracy, precision, recall, and F1 score, computed separately for each class and reported alongside confusion matrices. Since both datasets are balanced, these metrics provide a reliable and symmetric evaluation across classes.





\vspace{-15pt}
\section{Results}
\label{sec:results}
\vspace{-10pt}


We present the results of LLM-based annotation across the two tasks, evaluating the impact of prompt design and model architecture on classification performance. We additionally analyze the relationship between model errors and annotation ambiguity, using the human agreement distributions characterized in Section~\ref{sec:datasetecc} as a reference. Finally, we report the results of the ensemble strategy introduced in Section~\ref{sec:methodology} as an exploratory extension.

\vspace{-10pt}
\subsection{Labelling positive interactions}
\vspace{-5pt}

\begin{figure}[t]
    \centering
    \includegraphics[width=1.1\textwidth]{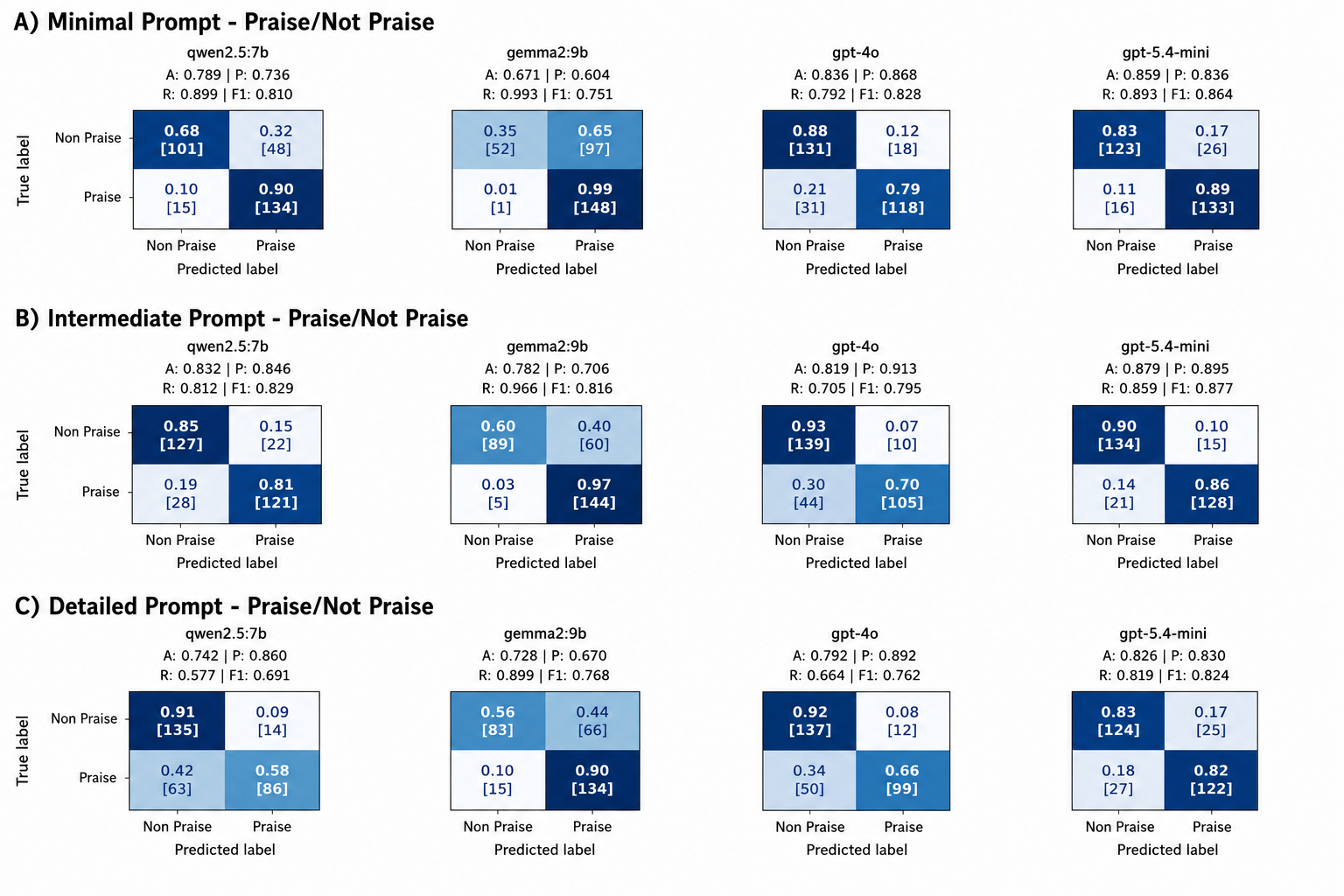} \vspace{-25pt}
    \caption{Confusion matrices for the praise classification task across four different models. Each cell reports the row-normalized rate (i.e., the proportion of instances in that true class predicted as each label) and, in brackets, the corresponding raw count. Above each matrix, we report the corresponding Accuracy, Precision, Recall, F1.} \label{fig:cm_praise} \vspace{-15pt}
\end{figure}

We first analyze the performance of LLMs on the praise classification task. Overall, we can observe that model performance varies depending on both the model architecture and the prompt design.

The intermediate prompt yields the best results across models, as shown in Figure~\ref{fig:cm_praise} panel B, outperforming both the minimal and structured configurations (Figure~\ref{fig:cm_praise}, panels A and C). The minimal prompt appears to provide insufficient guidance for identifying person-directed expressions, while the structured prompt, despite its additional instructions and examples, does not yield further gains, suggesting that excessive prompt complexity may introduce noise or reduce generalization, particularly for subtle or context-dependent interactions. Looking at the comparison between models, GPT-5.4-mini achieves the best overall results across all three types of prompts. This is likely due to its larger size and more extensive training, which enable it to better capture subtle linguistic cues and complex interaction patterns.
Notably, however, open-weight models delivered competitive results on this task. In particular, Qwen2.5 (7B) performs well when paired with the intermediate prompt, representing a viable alternative despite its small size. This suggests that prompt design can partially compensate for differences in model capacity. This is practically relevant, as open-weight models offer additional advantages in terms of accessibility, deployment cost, and the ability to run locally without relying on external APIs.



To situate these results with respect to task difficulty, we examine the distribution of model errors across instances with different levels of annotator agreement. In particular, we focus on instances which represent the most ambiguous cases in the dataset.
As shown in Table \ref{tabpraise}, a substantial portion of model errors occurs in these borderline cases across some models. For example, GPT-4o makes 27 out of 54 total errors (50\%) on comments where the ground truth label is determined by a minimal majority (4/7 annotator), while GPT-5.4-mini, despite achieving the lowest number of total errors (36), still produces 17 errors in these questionable instances (46\%).
This concentration of errors on low-agreement instances suggests that model limitations on this task are closely tied to the intrinsic ambiguity of praise expressions, consistent with the annotation complexity observed in Section~\ref{sec:datasetecc}.


\setlength{\tabcolsep}{12pt}
\begin{table}[t]
    \centering
    \caption{Distribution of model errors on the praise dataset, showing total errors vs errors on human low-agreement instances.}\label{tabpraise}
    \resizebox{0.6\textwidth}{!}{%
    \begin{tabular}{@{}lcc@{}}
    \toprule
    \textbf{Model} &  \textbf{Total Errors} & \textbf{Low-Agreement Cases}\\
    \midrule
    Qwen2.5:7b &  50 & 19\\
    Gemma2:9b &  65 & 15\\
    GPT-4o & 54 & 27\\
    GPT-5.4-mini & 36 & 17\\
    \bottomrule
    \end{tabular}} \vspace{-10pt}
\end{table}

\vspace{-10pt}
\subsection{Labelling negative interactions}
\vspace{-5pt}

\begin{figure}[t]
    \centering
    \includegraphics[width=1.1\textwidth]{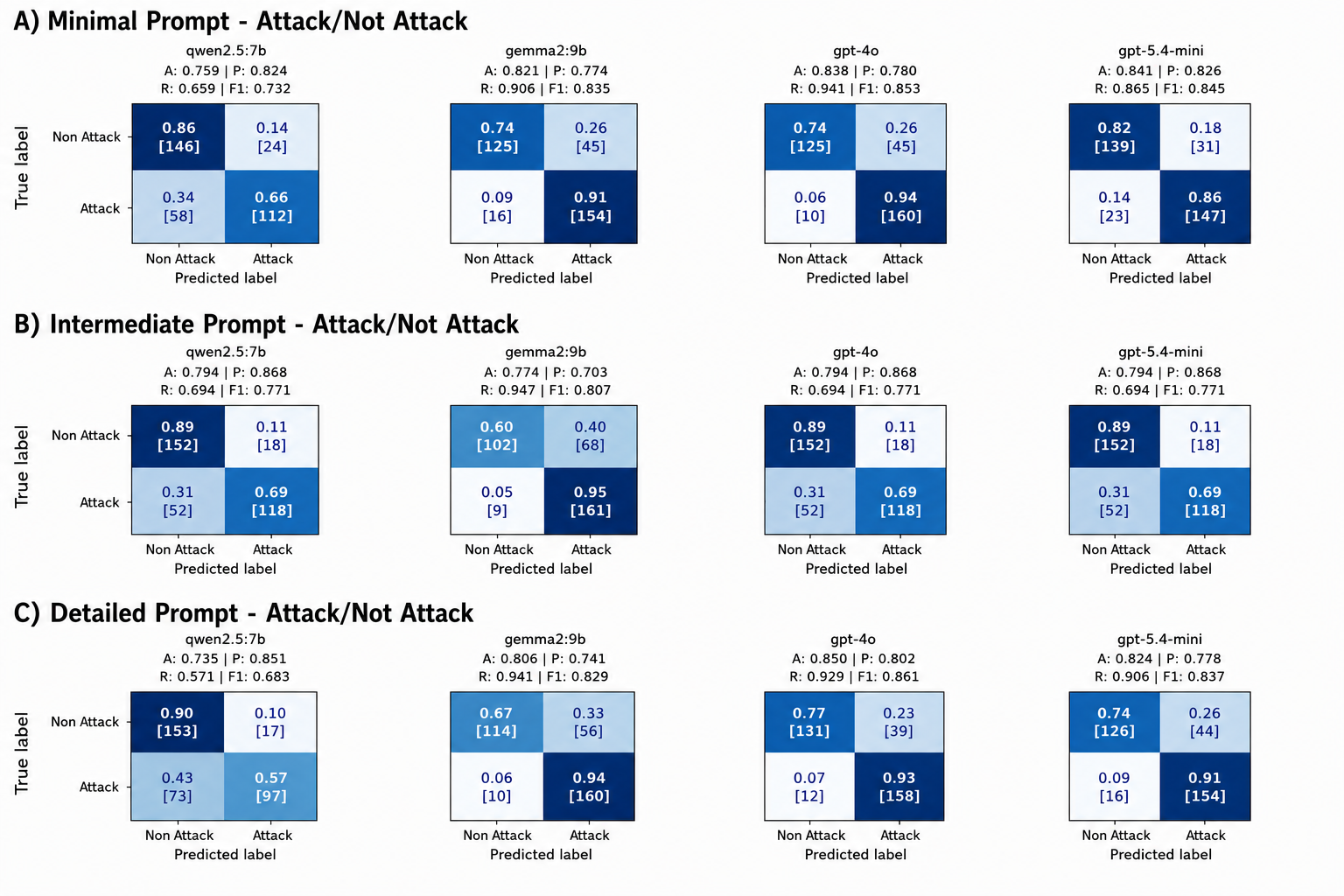} \vspace{-25pt}
    \caption{Confusion matrices for the attack classification task across four different models. Each cell reports the row-normalized rate (i.e., the proportion of instances in that true class predicted as each label) and, in brackets, the corresponding raw count. Above each matrix, we report the corresponding Accuracy, Precision, Recall, F1.}
    \label{fig:cm_attack} \vspace{-15pt}
\end{figure}

Results on the attack task reveal a markedly different pattern with respect to prompt design. Here, the minimal prompt yields the best performance across models (Figure~\ref{fig:cm_attack} panel A), outperforming both the intermediate and structured variants (Figure~\ref{fig:cm_attack} panels B and C). This contrasts with the praise task and suggests that personal attacks are expressed through sufficiently explicit and recognizable linguistic patterns that minimal guidance is enough,  and that additional prompt constraints may instead limit the model's ability to generalize across varied inputs.
Unlike in the praise task, open-weight models do not achieve performance comparable to proprietary models here. While they remain fairly stable across prompt variants, their results are consistently below those of GPT-4o and GPT-5.4-mini, indicating that model capacity and pre-training play a more decisive role than prompt design for attack detection, a gap that appropriate prompting cannot bridge as effectively as in the praise case.

\noindent


The error analysis tells a different story from the praise task. As shown in Table~\ref{tabattack}, while some errors do fall on borderline instances (those with 3/5 annotator agreement), the proportion is more evenly distributed across the dataset than in the praise case. This suggests that model errors on attack detection are not strongly tied to annotation uncertainty, consistent with the higher inter-annotator agreement observed for this task in Section~\ref{sec:datasetecc}. Attack detection thus presents challenges of a different nature, not primarily driven by subjective interpretation.


\begin{table}[t]
    \centering
    \caption{Distribution of model errors on the attack dataset, showing total errors vs errors on human low-agreement instances (3/5 annotator agreement).}
    \label{tabattack}
    \resizebox{0.6\textwidth}{!}{%
    \begin{tabular}{@{}l c c @{}}
    \toprule
    \textbf{Model} & \textbf{Total Errors} & \textbf{Low-Agreement Cases}\\
    \midrule
    Qwen2.5:7b & 82 & 25 \\
    Gemma2:9b & 61 & 21 \\
    GPT-4o & 55 & 17 \\
    GPT-5.4-mini & 54 & 19 \\
    \bottomrule
    \end{tabular}} \vspace{-15pt}
\end{table}

\vspace{-10pt}
\subsubsection{Testing ensembles of models}

\begin{figure}[t]
    \centering

    \begin{subfigure}[t]{0.32\textwidth}
        \centering
        \includegraphics[width=\textwidth]{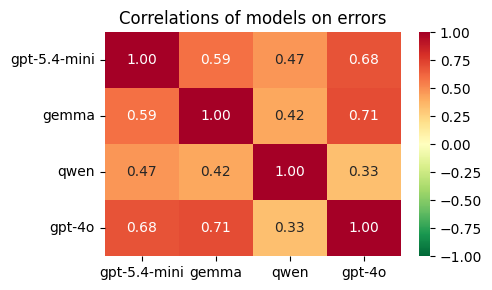}
        \caption{Error correlation on the negative full dataset.}
        \label{correlation_full}
    \end{subfigure}
    \hfill
    \begin{subfigure}[t]{0.32\textwidth}
        \centering
        \includegraphics[width=\textwidth]{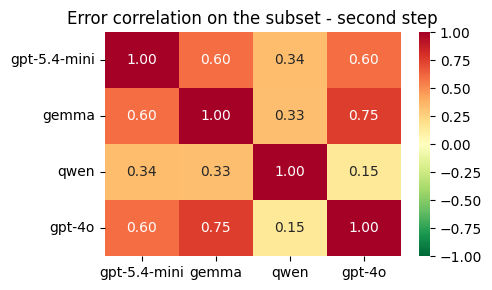}
        \caption{Error correlation on the subset predicted as attack by GPT-4o.}
        \label{correlation_step2}
    \end{subfigure}
    \hfill
    \begin{subfigure}[t]{0.32\textwidth}
        \centering
        \includegraphics[width=\textwidth]{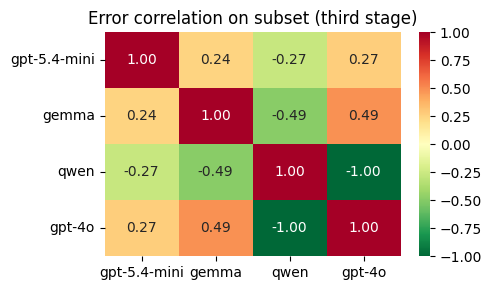}
        \caption{Error correlation on the subset predicted as attack by GPT-4o and not-attack by Qwen2.5:7b.}
        \label{correlation_step3}
    \end{subfigure}

    \caption{Error correlation matrices calculated to guide LLM choice in the pipeline.}
    \label{correlation}
    \vspace{-10pt}
\end{figure}

As an exploratory extension, we evaluate the cascading ensemble strategy described in Section~\ref{sec:methodology} on the attack classification task, under the best-performing configuration identified above (minimal prompt). Among all tested cascade configurations, the sequence GPT-4o $\rightarrow$ Qwen2.5:7b $\rightarrow$ GPT-5.4-mini yields a modest improvement over the best single-model result, reducing total errors from 54 (GPT-5.4-mini alone) to 49.
GPT-4o is selected as the primary model based on its low false positive rate (0.06 on the minimal prompt, Figure~\ref{fig:cm_attack}), meaning it is highly reliable at identifying non-attacks. Its predictions are therefore accepted for all instances it classifies as non-attack, while instances classified as attack (where GPT-4o is less certain) are passed to a second model. Qwen2.5:7b is selected for this second stage based on its low correlation of errors with GPT-4o on this subset (Figure~\ref{correlation}.b), meaning it tends to recover instances that GPT-4o misses. Finally, instances that Qwen2.5:7b classifies as non-attack (where it is less reliable) are passed to GPT-5.4-mini for a final refinement, selected for its low error correlation with \textit{both} previous models on the remaining uncertain cases (Figure~\ref{correlation}.c).

\begin{figure}[t]
    \centering
    \includegraphics[width=0.4\textwidth]{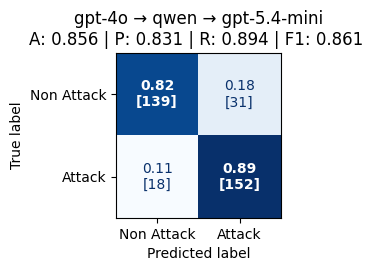}\vspace{-10pt}
    \caption{Confusion matrix of the cascaded ensemble (GPT-4o, Qwen2.5:7b, GPT-5.4-mini) on the negative interactions.}
    \label{fig:catena}
    \vspace{-15pt}
\end{figure}

The improvement obtained with the ensemble is not uniform across error types: recall increases through a reduction in false negatives, while false positives remain unchanged, as shown in Figure~\ref{fig:catena}. The ensemble is therefore effective at recovering missed attack instances but does not help in correcting over-prediction. Moreover, the same strategy applied to the praise task does not yield improvements (plot omitted due to lack of space). It is worth noting that the ensemble construction relies on error correlation matrices computed on held-out predictions with access to ground truth labels (information that would not be available in a real deployment scenario). The fact that even under these favorable, oracle-like conditions the improvement over the best single model is marginal (reducing total errors from 54 to 49) suggests that the cascading strategy does not offer a promising direction for this task. Overall, the results are inconclusive at best: the added complexity is unlikely to be justified in practice, and the conditions under which ensemble strategies could provide meaningful gains for relational annotation remain an open question.

\vspace{-15pt}
\section{Discussion}
\label{sec:discussion}
\vspace{-10pt}

The results presented in Section~\ref{sec:results} offer several insights into the use of LLMs as relational annotators for sign prediction, as well as into the nature of the two tasks themselves.

A first key finding is that LLMs in a zero-shot setting achieve good classification performance on both tasks, without any task-specific fine-tuning or labeled training data. This is encouraging for the broader goal of sign prediction in real-world settings, where annotated data is typically scarce and the cost of manual labeling is high. It suggests that LLM-based relational annotation is a practically viable approach, and that the methodology proposed in Section~\ref{sec:methodology} can be deployed with relatively modest resources.

Beyond overall performance, the results reveal that attack detection and praise detection behave differently along two dimensions: sensitivity to prompt design, and alignment between model errors and annotation ambiguity. Attack detection is largely insensitive to prompt complexity and model errors are not strongly concentrated on borderline instances, suggesting that hostile interactions carry sufficiently explicit linguistic signals that LLMs can identify reliably regardless of how the task is framed. Praise detection, by contrast, is sensitive to prompt design and model errors cluster on low-agreement instances, reflecting the greater context-dependence and subjectivity of positive relational gestures. This asymmetry has a direct implication for sign prediction: negative interactional signals can be extracted more directly and reliably, while positive ones require more careful prompt engineering and may remain inherently noisier.

A further finding concerns the role of model capacity. For praise detection, appropriate prompt design can partially compensate for differences in model size, with the open-weight Qwen2.5:7b achieving competitive performance under the intermediate prompt. For attack detection, this compensation is less effective, and proprietary models maintain a more pronounced advantage. This suggests that the two tasks place different demands on model capability, and that the choice of model should be guided by the specific annotation target rather than treated as a single decision.

\vspace{-10pt}
\paragraph{Limitations and open questions.} As with any empirical study of this kind, the present work leaves several questions open that point toward promising directions for future research. Both datasets are relatively small and constructed from specific sources; evaluating the methodology on larger and more diverse corpora would help establish how well the results generalize to other platforms and interaction styles. Additionally, as noted in Section~\ref{sec:datasetecc}, the observed difference in annotation complexity between the two tasks may partly reflect dataset-specific factors, and further investigation across multiple corpora would help disentangle these effects. The prompt designs evaluated here are predefined zero-shot formulations; exploring more flexible strategies such as few-shot prompting, chain-of-thought reasoning, or lightweight fine-tuning represents a natural avenue for improving performance further, particularly on the praise task. The ensemble strategy explored in Section~\ref{sec:results} yields only marginal gains on attack detection and does not generalize to praise detection, suggesting that the conditions under which model combination is beneficial for relational annotation merit further investigation. Finally, the most important open question is the integration of relational annotation into a downstream sign prediction pipeline: the ultimate impact of annotation quality on graph-level tasks (such as signed link prediction or community detection in signed networks) remains to be assessed, and constitutes the natural next step for this line of work.

\vspace{-15pt}
\section{Conclusion}
\label{sec:conclusion}
\vspace{-10pt}

This paper investigated whether large language models in a zero-shot setting can effectively support sign prediction in social networks by providing relational annotations of textual interactions: specifically, classifying personal praise and personal attacks as indicators of positive and negative social ties. The key motivation is to move beyond sentiment analysis, which conflates the valence of the content exchanged between users with the valence of the relationship itself, a distinction that prior approaches have struggled to operationalize.
Our results demonstrate that zero-shot LLMs achieve good classification performance on both tasks without any task-specific training data, establishing a practical and resource-efficient baseline for relational annotation. Performance does, however, depend on both the task and the model. Attack detection is robust across prompt variants and model types, reflecting the explicit nature of hostile linguistic signals. Praise detection is more sensitive to prompt design and more affected by the intrinsic subjectivity of the task, with model errors concentrating on instances where human annotators themselves disagree. More generally, these results suggest that LLM-based relational annotation is a viable and flexible building block for any sign prediction framework that operates by aggregating interaction-level signals into relationship-level polarities.

\vspace{-5pt}

\vspace{-10pt}

\bibliographystyle{splncs04}
\bibliography{biblio.bib}

\appendix

\vspace{-20pt}
\section{Details of prompts}
\label{app:prompts}
\vspace{-10pt}

In this section we report the prompts administered to the four models (Qwen2.5:7b, Gemma2:9b, GPT-4o, GPT-5.4-mini). The prompts are divided by type (Praise vs. Attack) and by level of complexity (minimal, intermediate, structured).

\vspace{-10pt}
\subsection{Prompts for praises}
\label{app:prompts_praise}
\vspace{-5pt}

\lstset{
    basicstyle=\scriptsize\ttfamily,
    breaklines=true, 
    frame=lines,     
    backgroundcolor=\color{gray!5},
    captionpos=b     
}

\begin{lstlisting}[caption={Praise: Minimal level}, label={prompt:minpraise}]
[You are a linguistic annotator. Label this sentence as praise directed at the interlocutor (1) or not (0).]
\end{lstlisting}\vspace{-10pt}

\begin{lstlisting}[caption={Praise: Intermediate level}, label={prompt:intpraise}]
[You are a linguistic annotator. Label this sentence as praise directed at the interlocutor (1) or not (0). Label 0 if the praise refers to a third party (he, she, they, or a name).]
\end{lstlisting}\vspace{-10pt}

\begin{lstlisting}[caption={Praise: Structured level}, label={prompt:detpraise}]
[You are analyzing online comments. For each comment, decide whether it is a person-directed praise (1) or not (0).
A comment is praise (1) if it expresses something positive toward the person being addressed, including:
- Direct compliments ("You are amazing", "Good job")
- Congratulations ("Congrats!", "Well done")
- Thanks combined with positive sentiment ("Thanks, looks great!")
- Indirect or minimal praise ("Nice catch", "Brave soul", "Clever!")
- Praise directed at something belonging to the person ("your work is beautiful")
A comment is NOT praise (0) if:
- It is neutral or purely informational
- It is positive about a topic but NOT toward the person being addressed
- It expresses something positive about a third party 
(someone not involved in the conversation)]
\end{lstlisting} \vspace{-10pt}

\vspace{-10pt}
\subsection{Prompts for attacks}
\label{app:prompts_attacks}
\vspace{-5pt}

\begin{lstlisting}[caption={Attack: Minimal level}, label={prompt:minattack}]
[You are analyzing online comments. Determine whether the comment is a direct personal attack (1) or not (0).]
\end{lstlisting}\vspace{-10pt}

\begin{lstlisting}[caption={Attack: Intermediate level}, label={prompt:intattack}]
[You are a linguistic annotator. Label this comment as a direct personal attack directed at the interlocutor (1) or not (0). Label 0 if the attack refers to a third party (he, she, they, or a name) or if there is no attack at all.]
\end{lstlisting}\vspace{-10pt}

\begin{lstlisting}[caption={Attack: Structured level}, label={prompt:detattack}]
[You are analyzing online comments. For each comment, decide whether it is a person-directed attack (1) or not (0).
A comment is an attack (1) if it expresses something negative toward the person being addressed, including:
- Direct insults ("You are stupid", "You're an idiot")
- Offensive or demeaning language ("This is a dumb take", "You're clueless")
- Mockery or sarcasm targeting the person ("Nice job genius", "Brilliant, as always" used sarcastically)
- Aggressive or hostile tone directed at the person ("Shut up", "Get lost")
- Negative statements about the persons abilities, character, or behavior ("You have no idea what you're doing")
A comment is NOT an attack (0) if:
- It is critical but not directed at the person (e.g., criticism of ideas, topics, or content)
- It expresses disagreement in a respectful or neutral way
- It discusses negative topics without targeting the person
- It is negative toward a third party (someone not involved in the conversation).]
\end{lstlisting}\vspace{-10pt}

\end{document}